\documentclass[twocolumn,floatfix,pra,aps,showpacs]{revtex4}
\usepackage{epsfig,graphicx}
\usepackage{flafter}
\usepackage{amsmath}
\usepackage{color}
\usepackage{braket}
\newcommand{\be}{\begin{equation}}
\newcommand{\ee}{\end{equation}}

\begin{document}

\title{Quadrupole oscillation  in a dipolar Fermi gas: hydrodynamic vs collisionless regime}
\author{M. Abad}
\affiliation{Departament d'Estructura i Constituents de la Mat\`{e}ria,\\
Facultat de F\'{\i}sica, Universitat de Barcelona, E--08028 Barcelona, Spain}
\author{A. Recati}
\affiliation{INO-CNR BEC Center and Dipartimento di Fisica, Universit\`a di Trento, 38123 Povo, Italy}
\author{S. Stringari}
\affiliation{INO-CNR BEC Center and Dipartimento di Fisica, Universit\`a di Trento, 38123 Povo, Italy}

\begin{abstract}
The surface quadrupole mode of an harmonically trapped  dipolar Fermi gas is studied in both the hydrodynamic and  collisionless  regimes.  The anisotropy and long range effects of the dipolar force as well as the role of the trapping geometry are explicitly investigated.
In the hydrodynamic regime the frequency is always slightly  smaller than the  $\sqrt{2}\omega_\perp$ value holding for gases interacting with contact interactions. In the collisionless regime the frequency can be either pretty smaller or larger than the non-interacting value $2\omega_\perp$, depending on the cloud aspect ratio. 
Our  results suggest that  the frequency of the surface quadrupole oscillation can provide a useful test for studying, at very low temperatures,  the transition between the normal and the superfluid phase and, in the normal phase at higher temperatures,   the crossover  between the collisional and collisionless regimes.  The consequences of the anisotropy  of the dipolar force on the virial theorem are also discussed. 
\end{abstract}

\maketitle

\section{Introduction}

Soon after the  experimental realization of Bose-Einstein condensation and  Fermi superfluidity in atomic gases, much attention has been devoted to the study of collective oscillations  (for reviews  see, for example, \cite{rmp1,rmp2}). Collective oscillations actually  provide unique and valuable information on the macroscopic and dynamic behavior of  these  quantum gases, being  sensitive to both the equation of state,  the effects of superfluidity  and the role of collisions. While compressional modes are directly sensitive to the equation of state and have been employed to test  its fine details with high precision (see, for example, the case of  the Fermi gas along the BEC-BCS crossover \cite{grimm}), surface oscillations are more suited
 to understand whether the gas is in the hydrodynamic or collisionless regime. This is especially important in view of the possibility of identifying  the superfluid 
phase of the gas, where the dynamics of the low frequency modes is always governed by the equations of irrotational hydrodynamics. 

The recent experimental availability of atomic gases interacting with magnetic dipolar forces \cite{pfau} and the very recent progress in the realization of ultracold heteronuclear molecules interacting with electric dipolar forces \cite{Jin} have  already stimulated a significant number of theoretical papers (for a review see, for example, \cite{dipoleReview}).  Special attention has been also focused on the study of the collective oscillations of both Bose \cite{collBose} and Fermi \cite{collFermi2,collFermi,Lima2010a,Lima2010} dipolar gases, including the study of long range effects in bi-layer systems \cite{bilayers}.   A major focus of most of these  papers was the problem of the instability caused by the attractive part of the dipolar force which shows up in the appearence of  an imaginary value of the collective frequency of  compressional modes. In this paper we will instead focus on the  main features of the surface oscillations with the aim to explore the consequences of the anisotropic and long range nature of the dipolar force in both the hydrodynamic and collisionless regimes. In this work the dipolar effects are taken into account within mean-field theory.

For harmonically trapped  gases interacting with zero range forces, which is the case of  atomic gases without dipolar interactions, the  frequency of the surface quadrupole mode, in the hydrodynamic regime, is given by the universal value $\sqrt 2 \omega_\perp$ \cite{stringari96}
where  $\omega_\perp$ is the radial oscillator frequency  in the $x-y$ plane where we  excite the collective oscillation. This value differs from the non-interacting value  $ 2 \omega_\perp$ of the ideal gas since in the hydrodynamic regime the kinetic energy does not contribute to the restoring force of the quadrupole oscillation which is then entirely provided by the external harmonic potential. In the collisionless regime the kinetic energy does instead contribute to the restoring force and  the  value of the frequency of the quadrupole oscillation depends on the explicit balance between the kinetic  and external potential contributions. This balance is affected by the presence of two-body interactions and is usually accounted for by the virial relationship \cite{vichi}. The anistropic and long range nature of the dipolar force affects the above scenario both in the hydrodynamic  and in the collisionless regimes.  We will mainly focus on the case of Fermi dipolar gases where the quest of superfluidity is more challenging for single spin species gases and the competition between the hydrodynamic and collisionless regimes is relevant also at  very low temperatures.  The paper is organized as follows. In Sect. II we discuss the equilibrium properties of the gas and the anisotropy effects of the dipolar interaction on the density profile. We furthermore develop a variational approach, based on  a scaling ansatz, to derive the equations of motion characterizing the quadrupole oscillation in both the hydrodynamic and collisionless regimes. In Sect. III we discuss the resulting effects on the collective frequencies and the temperature conditions for being in the collisionless or hydrodynamic regime. In Sect. IV 
we provide the explicit derivation of the  virial theorem and we point out its relevance for experiments where one suddenly switches off the dipolar interaction. In Sect. V we draw our conclusions, while in the Appendix A we present the results for  a few  relevant dimensionless functions whose knoweldge is needed   for the actual calculation of the collective frequencies. 

\section{Trapped dipolar Fermi gases}
We consider a Fermi gas of dipolar atoms or molecules, confined in an axially symmetric harmonic trap
\begin{equation}
	V_{\text{ho}}(\mathbf{r})=\frac{1}{2}m(\omega_\perp^2r_\perp^2 + \omega_z^2 z^2)\,,
\end{equation}
where $m$ is the mass of a single fermion, $\omega_\perp$ and $\omega_z$ are the radial and axial trapping frequencies, respectively,  and $r^2_\perp\equiv x^2+y^2$. If the dipoles are polarized along $z$, the microscopic dipolar interaction can be written as
\begin{equation}
	v_{dd}(\mathbf{r}-\mathbf{r^\prime})=\frac{d^2 }{|\mathbf{r}-\mathbf{r^\prime}|^3}\left(1-3\cos^2\theta\right)\, ,
\end{equation}
with $d^2=p^2/4\pi\epsilon_0$ for electric dipoles ($p$ is the electric dipole moment)
$d^2=\mu_0\mu^2/4\pi$ for magnetic dipoles ($\mu$ is the magnetic dipole moment)
The angle $\theta$ is the angle that the polarization axis forms with the relative distance between two dipoles.

\subsection{Equilibrium properties}
\label{ground}

The ground state properties of the system can be calculated employing a variational approach starting from the  following energy functional, based on a mean field picture:
\begin{equation}
	E=\braket{\Psi|H|\Psi} =  E_{\text{kin}}+ E_{\text{ho}} + E_{\text{dip}}
\label{energy}
\end{equation}
where
\begin{align}
	&E_{\text{kin}} =\frac{3}{5}\frac{\hbar^2}{2m}(6\pi^2)^{2/3} \int d{\bf r}[n({\bf r})]^{5/3}\label{EkTF}\\
	&E_{\text{ho}} = \int d{\bf r} n({\bf r})V_{\text{ho}}\\
	&E_{\text{dip}}=\int d{\bf r}_1d{\bf r}_2 n({\bf r}_1) n({\bf r}_2) v_{dd}({\bf r}_1-{\bf r}_2) 
\end{align}
are, respectively, the kinetic, oscillator and dipolar energies.
In writing the above equations we have used a Thomas-Fermi description for the kinetic energy, under the assumption of isotropy of the momentum distribution. This assumption, which  applies to  three-dimensional configurations, neglects possible anisotropy effects caused by the  dipole interaction in momentum space. In the expression for the dipole energy we have further ignored the occurrence of exchange terms.  In \cite{Miyakawa2008, collFermi, Lima2010} it was shown that both the anisotropy of the kinetic energy and the exchange effect in the dipole energy can be relevant for establishing the exact conditions of stability of the gas, but are negligible if one considers configurations sufficiently far from the instability point. Notice the absence, in Eq.(\ref{energy}), of the contact term fixed by the s-wave scattering length which typically enters the  energy functional of interacting  Bose gases. This term is absent in the single species Fermi gas considered in the present paper.  

The equilibrium density profile of the trapped gas is obtained by imposing minimization of the energy $E$ with  the proper normalization constraint, which is obtained adding the grand canonical term $-\mu\int d{\bf r}\,n({\bf r})$ to the energy functional, with $\mu$ the chemical potential. In general the equilibrium profile will be a function of the quantity $(\mu - V_{\text{ho}}({\bf r}))$. In the following, for simplicity, we will make the further assumption that the density profile takes the  form 
\begin{equation}
	n(\mathbf{r}) =\frac{8}{\pi^2}\frac{N}{R_\perp^2R_z}\left(1-\frac{r_\perp^2}{R_\perp^2} - \frac{z^2}{R_z^2}\right)^{3/2}\label{TF}
\end{equation}
where $N$ is the number of fermions and  the radii $R_\perp$ and $R_z$ are  variational parameters. The density Eq.(\ref{TF}) would correspond to the exact Thomas-Fermi solution of  a Fermi gas  in the absence of the dipolar interaction term. 
Using this  ansatz, the dipole energy $E_{\text{dip}}$ can be evaulated analytically and takes the form
\cite{Lima2010,collFermi}: 
\begin{align}
	&E_{\text{dip}}=-\frac{1}{4}N\epsilon_F\varepsilon_{dd} f(\kappa). \label{edip} 
\end{align}
In the previous expression $\epsilon_F=\hbar^2 k_F^2/2m$ is the Fermi energy,  $k_F$  being the Fermi momentum defined by the normalization condition $R_\perp^2R_zk_F^3=48N$ and we have introduced the 
dimensionless parameter 
\begin{equation}
	\varepsilon_{dd}=\frac{2^{11}}{3^4 35\pi^2}\frac{d^2(k_F)^3}{\epsilon_F} .
\end{equation} 
The dimensionless function $f(\kappa)$ is defined in the Appendix, with $\kappa=R_\perp/R_z$ the aspect ratio of the trapped cloud \cite{notefkappa}.
 Notice that in general $\kappa$ differs from the aspect ratio $\lambda=\omega_z/\omega_\perp$ of the trapping potential as a consequence of the anisotropy of the dipolar interaction.
It is also interesting to notice that the term in front of  $f(\kappa)$ in the dipolar energy Eq.(\ref{edip}) scales like $1/R_xR_yR_z$, which is the same scaling dependence of a zero-range contact force.  Deviations from the behaviour of a contact force, caused by the anisotropy and finite-range nature of the dipolar force,  are then entirely accounted for  by the  function $f(\kappa)$.

The equilibrium configuration is obtained by imposing the stationarity of the energy functional Eq.(\ref{energy}) with respect to variations of the Thomas-Fermi radii. One then finds the following equations:
\begin{align}
	&\frac{R_\perp}{l_\perp} =k_F l_\perp \sqrt{1- \varepsilon_{dd}\left[1-\frac{3}{2}\frac{\kappa^2}{1-\kappa^2}f(\kappa)\right]}\label{Rperp}\\
	&\frac{R_z}{l_z} = k_F l_z \sqrt{1 + 2\varepsilon_{dd}\left[1-\frac{3}{2}\frac{1}{1-\kappa^2}f(\kappa)\right]}\label{Rz}
\end{align}
where $l_i=\sqrt{\hbar/m\omega_i}$, $i=\perp,z$ is the harmonic oscillator length in the $i$-th direction. Combining equations (\ref{Rperp}) and (\ref{Rz})  one finds a transcendental equation for $\kappa$ as a function of $\lambda$ and $\varepsilon_{dd}$,
\begin{equation}
	\frac{\kappa^2}{\lambda^2}\left[3\varepsilon_{dd}\frac{f(\kappa)}{1-\kappa^2}\left(\frac{\lambda}{2}+1\right)-2\varepsilon_{dd}-1\right]=\varepsilon_{dd}-1\label{Trans}
\end{equation}
Note that this expression is the same as the one found for bosons \cite{ODell2004}, provided the parameter $\varepsilon_{dd}$ is properly defined. 

\begin{figure}
\centering
 \epsfig{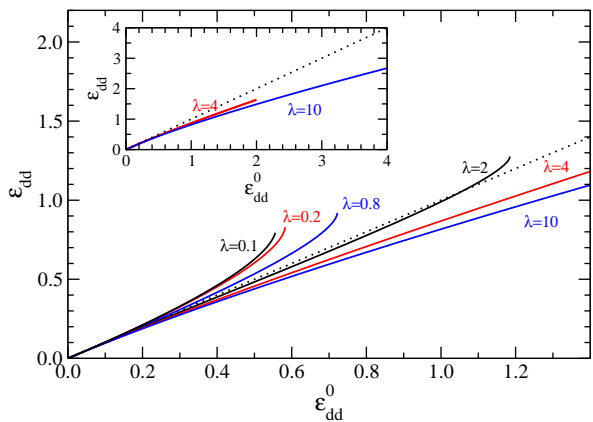}
\caption{Density-dependent dipolar parameter $\varepsilon_{dd}$ as a function of the externally controllable dipolar paramenter $\varepsilon_{dd}^0$.}\label{edd0vedd}
\end{figure}

It is useful to express the previous quantities in terms of experimentally controllable parameters. 
We define the non-interacting Fermi energy $\epsilon_F^0=\hbar(6N\omega_\perp^2\omega_z)^{1/3}$, 
the non-interacting Fermi momentum $k_{F}^0=\sqrt{2 m \epsilon_F^0}/\hbar$ and the corresponding dimensionless parameter
\begin{equation}
	\varepsilon_{dd}^0=\frac{2^{11}}{3^4 35\pi^2}\frac{d^2(k_F^0)^3}{\epsilon_F^0}
\end{equation} 
In Figure \ref{edd0vedd} we show the relationship  between the relevant interaction parameter $\varepsilon_{dd}$ and $\varepsilon_{dd}^0$. Since $\varepsilon_{dd}/\varepsilon_{dd}^0=k_F/k_F^0$, the difference between the two quantities is due to the departure of $k_F$ from the non-interacting value $k_F^0$, caused by the dipolar interaction.  

\begin{figure}
\centering
\epsfig{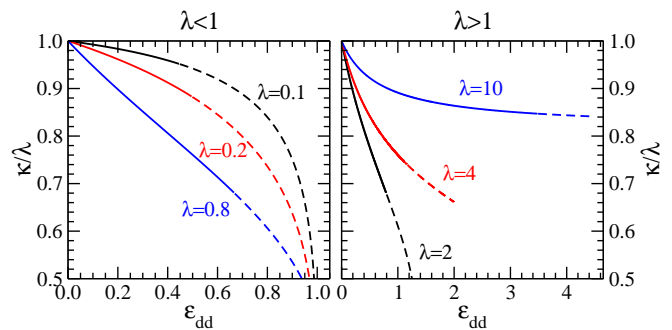}
\caption{Gas aspect ratio normalized by the trap aspect ratio, $\kappa/\lambda$, as a function of $\varepsilon_{dd}$. Left panel: for a cigar-shaped trapping potential; Right panel: for a pancake-shaped trapping potential. Dashed line marks instability as predicted by including exchange interactions (see for instance \cite{Miyakawa2008}).}\label{kolvedd}
\end{figure}

In Figure \ref{kolvedd} we report instead the dependence of $\kappa/\lambda$ on $\varepsilon_{dd}$ for different values of $\lambda$. This figure explicitly reveals how the dipolar interaction modifies the aspect ratio of the trapped gas with respect to the trap value $\lambda$.  For zero dipolar interactions we recover the non-interacting limit, for which $\kappa=\lambda$. Note also that, as expected, the magnetostriction (electrostriction) effect is larger for $\lambda\sim 1$, while it is much smaller for very small or large trapping aspect ratios. The dashed lines in the figure mark the regions for which the solution, including the exchange interactions, would  predict instability \cite{Miyakawa2008} (see also \cite{Lima2010,notenumerics}) and we will keep this notation all over the work.

\subsection{Variational approach to the $m=2$ surface quadrupole oscillation}
 
In the following we study the effect of the long-range dipolar interaction on the frequency of the  transverse ($m=2$)  quadrupole mode  assuming the gas is either in the collisionless (CL) or in the hydrodynamic (HD) regime. The results for the latter are always attained in the superfluid phase,  whose description is based on the equations of irrotational hydrodynamics.

In order calculate the collective mode frequencies we implement a variational approach based on the action   
\begin{equation}
	S = \int dt \Braket{\Psi | H - i\hbar\frac{\partial}{\partial t} | \Psi} \label{action}
\end{equation}
from which we can derive the equations of motion for configurations perturbed from equilibrium and hence  evolving in time. The $m=2$ quadrupole oscillation can be described employing a variational approach based on the  scaling transformation
\begin{equation}
	\ket{\Psi} = e^{i\alpha(t)\sum_k(x_k^2-y_k^2)}e^{ib(t)\sum_k ((x_kp^x_k-y_kp^y_k)+hc)}\ket{\Psi}_0
\end{equation}
 where the two time dependent parameters $\alpha(t)$ and $b(t)$ account, respectively,  for the presence of current terms and changes in the density distribution  with respect to the unperturbed configuration $\ket{\Psi}_0$.  Here $p_k^i$ is the $i-th$ component of the momentum and $k=1,\dots,N$ is the particle index.  The density distribution and the velocity field  associated with the wave function $\ket{\Psi}$ are given by
\begin{equation}
n(x,y,z)= n_0(e^{-b(t)}x,e^{b(t)}y,z)
\label{densscal}
\end{equation}
and 
\begin{equation}
{\bf v}= \frac{\hbar}{m}\alpha(t)\nabla(x^2-y^2)
\label{velfield}
\end{equation}
where $n_0$ is the equilibrium density profile.

The first order variation with respect to $\alpha$ and $b$ has to be zero, while from the second order variation we find the equations of motion for the parameters  $\alpha$ and $b$ in the linear regime from which we calculate the collective frequencies.  
The term associated with the time derivative in \eqref{action} gives:
\begin{equation}
	\Braket{\Psi|-i\hbar\frac{\partial}{\partial t}|\Psi} = 2\hbar\dot{\alpha}b N\langle r_\perp^2 \rangle \label{actiont}
\end{equation}
where the radial square radius $\langle r_\perp^2 \rangle$ should be calculated at equilibrium. 

The hamiltonian term in the action integral \eqref{action} is just the energy functional Eq. (\ref{energy}). However, out of equilibrium the expression for the kinetic energy differs from  the  Thomas-Fermi expression Eq. (\ref{EkTF}). First of all the presence of the velocity field Eq. (\ref{velfield}), caused by the scaling transformation, gives rise to a ``classical" contribution to the kinetic energy of the form
\begin{equation}
\delta E_{\text{kin}}(\alpha) = \frac{m}{2}\int d{\bf r}{\bf v}^2n({\bf r})=  \frac{2\hbar^2}{m}N\alpha^2\langle r_\perp^2\rangle .
\label{velcont}
\end{equation}
Second one should distinguish whether the system is in the hydrodynamic regime, where local equilibrium is restored during the oscillation, or in the collisionless regime. In the first case one can still use the Thomas-Fermi expression Eq. (\ref{EkTF}). Since this expression includes a local dependence on the density, the density change Eq. (\ref{densscal}) caused by the quadrupole scaling  transformation does not modify  the value of the kinetic energy which is consequently affected only by the velocity field contribution Eq. (\ref{velcont}). In the collisionless case the kinetic energy gets an additional  contribution due to the static part of the transformation  (term in $b$). The latter causes a deformation of the Fermi sphere in momentum space giving rise to an elastic type contribution, a  typical feature of Fermi systems \cite{stringari}. The total contribution to the changes in the kinetic energy term takes in this case the form
\begin{equation}
\delta E^{\text{(CL)}}_{\text{kin}} =\delta E_{\text{kin}}(\alpha)+2b^2E_{\text{kin},\perp},
\end{equation}
where $E_{\text{kin},\perp}$ is the radial contribution to the ground state kinetic energy. 

The other terms in the energy functional (oscillator and dipolar energy) depend only on the local density distribution and are consequently affected by the scaling transformation  through the term in $b(t)$: 
\begin{align}	
	&\delta E_{\text{ho}} (b)=   m\omega^2_\perp N b^2\langle r_\perp^2\rangle,\\
	&\delta E_{\text{dip}}(b)= - \frac{1}{8}b^2N\epsilon_F\varepsilon_{dd}g(\kappa).\label{edipaction}
\end{align}
In order to derive  the dipolar contribution Eq. (\ref{edipaction}) we have expanded the dipolar energy Eq. (\ref{edip}) to second order in $b$, using the scaling transformation $R_x\rightarrow e^{b}R_x$ and $R_y\rightarrow e^{-b}R_y$ and have introduced the function $g(\kappa)$ proportional to the second derivative of $E_{\text{dip}}(b)$ with respect to $b$ (see the Appendix A). 

Using Eqs. \eqref{actiont} and \eqref{velcont}--\eqref{edipaction} the action in the hydrodynamic regime reads
\begin{align}
	S^{(\text{HD})}  =&S^{(0)}+ \int dt \left[ \frac{2\hbar^2}{m}N\alpha^2\langle r_\perp^2\rangle +  m\omega^2_\perp N b^2\langle r_\perp^2 \rangle + \right.\nonumber\\
	&+2\hbar\dot{\alpha}b N\langle r_\perp^2 \rangle  - \frac{1}{8}b^2N\epsilon_F\varepsilon_{dd}g(\kappa)   \bigg],\label{varactHD}
	\end{align}
where $S^{(0)}$ is the action calculated for the ground state. In the collisionless regime one should add the term $\int dt 2b^2E_{\text{kin},\perp}$ to the right hand side of Eq. (\ref{varactHD}).  

Imposing that the action be stationary with respect to variations of $\alpha(t)$ and $b(t)$
the equations of motion can be written as
\begin{align}
	& 2\frac{2\hbar^2}{m}N\alpha \langle r_\perp^2\rangle -  2\hbar\dot{b} N\langle r_\perp^2 \rangle = 0 \\
	&  2 m\omega^2_\perp N b\langle r_\perp^2 \rangle + 2\hbar\dot{\alpha} N\langle r_\perp^2 \rangle  - \frac{1}{4}bN\epsilon_F\varepsilon_{dd}g(\kappa)   =0 \label{coupled2}
\end{align}
in the  hydrodynamic regime while, in the collisionless regime, they read
\begin{align}
	& 2\frac{2\hbar^2}{m}N\alpha \langle r_\perp^2\rangle -  2\hbar\dot{b} N\langle r_\perp^2 \rangle = 0 \\
	&  4bE_{\text{kin},\perp} + 2 m\omega^2_\perp N b\langle r_\perp^2 \rangle +  2\hbar\dot{\alpha} N\langle r_\perp^2 \rangle  - \frac{1}{4}bN\epsilon_F\varepsilon_{dd}g(\kappa)   =0 \label{coupled2}
\end{align}
The above equations admit harmonic  solutions.
 The frequency of the oscillation in the hydrodynamic regime is given by
\begin{equation}
	\omega_Q^{\text{(HD)}}=\sqrt{2}\omega_\perp\left[1-\frac{1}{4}\,\frac{\varepsilon_{dd}}{1+\varepsilon_{dd}\left(\frac{3}{2}\frac{\kappa^2f(\kappa)}{1-\kappa^2}-1\right)}\,g(\kappa)\right]^{1/2},\label{wQHD}
\end{equation}
which reduces to the usual hydrodynamic result $\sqrt{2}\omega_\perp$ \cite{stringari96} by setting $\varepsilon_{dd}$ equal to zero (notice, however, that in a single species Fermi gas the $\varepsilon_{dd}=0$ limit is incompatible with the  hydrodynamic regime which always requires the presence of interaction terms).   The same expression Eq. (\ref{wQHD}) for the quadrupole frequency in the hydrodynamic regime  holds also for bosons, where $\varepsilon_{dd}=md^2/3\hbar^2 a_s$ with $a_s$ the $s$-wave scattering length. 

In order to provide a compact expression for the quadrupole frequency in the collisionless regime it is useful to employ the radial virial relationship (see derivation in Sect. (\ref{sec:virial}))
\begin{equation}
2E_{\text{kin},\perp} - 2E_{\text{ho},\perp} + 2E_{\text{dip}} -\frac{1}{4}N\epsilon_F\varepsilon_{dd}h(\kappa)=0\label{virial}
\end{equation}
to express the kinetic energy contribution entering the equation of motion (\ref{coupled2}) 
in terms of the oscillator and dipolar energy contributions. Here $E_{\text{ho},\perp}=m\omega^2_\perp N b\langle r_\perp^2 \rangle/2$ and $E_{\text{dip}}$ is given by Eq. (\ref{edip}). The function $h(\kappa)=-\kappa df(\kappa)/d\kappa$ is proportional to the  first derivative of the function $f(\kappa)$ entering Eq. (\ref{edip}) and is explicitly calculated in the Appendix A.
Using the virial relationship (\ref{virial}) one can  derive the following useful expression for the quadrupole frequency in the collisionless regime: 
\begin{align}
	\omega_Q^{\text{(CL)}}=&2\omega_\perp\left[1+\frac{1}{2}\,\frac{\varepsilon_{dd}}{1+\varepsilon_{dd}\left(\frac{3}{2}\frac{\kappa^2f(\kappa)}{1-\kappa^2}-1\right)}\times\right.\nonumber\\
	&\times\left(f(\kappa)+\frac{1}{2}h(\kappa)-\frac{1}{4}g(\kappa)\right)\Bigg]^{1/2} \label{wQCL}
\end{align}
 showing that, as  $\varepsilon_{dd}\rightarrow0$, the non-interacting limit $2\omega_\perp$ is recovered.

Eqs. (\ref{wQHD}) and (\ref{wQCL}) represent the main result of the present paper. They give explicit predictions for the quadrupole frequencies in the hydrodynamic and collisionless regimes in terms of the relevant dimensionless  parameters $\varepsilon_{dd}$ and $\kappa$. These parameters are directly related to the experimentally controllable parameters $\varepsilon_{dd}^0$ and $\lambda$, as shown in Figs.  \ref{edd0vedd} and \ref{kolvedd}. 

\section{Quadrupole mode in the hydrodynamic and collisionless regimes}

We are interested in comparing the frequency of the  radial quadrupole mode in  the hydrodynamic and  collisionless regimes.
Let us first analyze Eqs.~(\ref{wQHD}) and (\ref{wQCL}) separately. 
In the hydrodynamic  regime the quadrupole frequency is shifted with respect to the usual hydrodynamic result $\sqrt 2\omega_\perp$, holding for gases interacting with  contact forces. The deviations are caused by the anisotropic and finite range  character of the dipolar interaction. This shift vanishes for $\kappa\rightarrow0$ and $\kappa\rightarrow\infty$, since in these two limits the system is so deformed that the dipole force effectively  behaves like a contact force.  This is reflected in the flat behavior of $f(\kappa)$ for very pancake and very cigar-shaped traps, as well as in the fact that $g(\kappa)$ tends to zero in these limits (see the Appendix A).
The top panels of Fig.~\ref{wq} show the behavior of the quadrupole frequency in the HD regime as a function of the dipolar parameter $\varepsilon_{dd}$ for different trap aspect ratios $\lambda$. The dashed lines mark the values of  $\varepsilon_{dd}$  for which the Fermi gas is no longer stable, according to \cite{collFermi}. The figure shows that the frequency is always smaller than the $\sqrt2\omega_\perp$ value and that the shift is largest for intermediate trap aspect ratios.  However, as already pointed out in Ref.~\cite{Lima2010}, the shift of the surface quadrupole oscillation is in general small compared to the case of compressional modes which are much more sensitive to the equation of state.

\begin{figure}
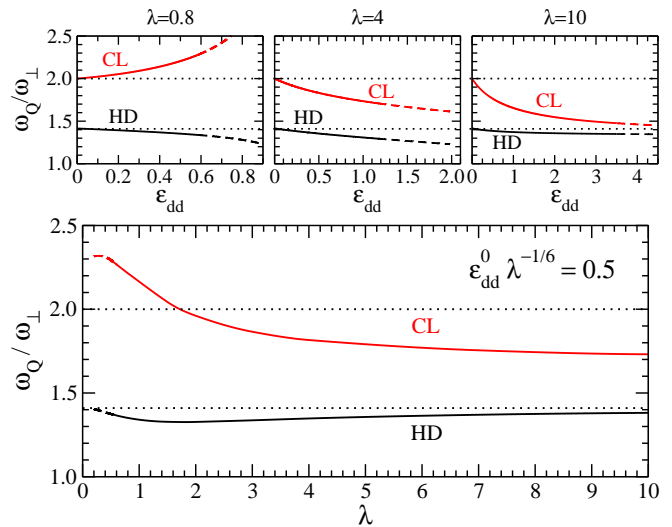

	\epsfig{file=wqvedd.eps, width=\linewidth, clip=true}
	\epsfig{file=wqvl.eps, width=\linewidth, clip=true}
\caption{Top panels: Quadrupole frequency as a function of the dipolar strength $\varepsilon_{dd}$, both for the CL and the HD regimes and for different trap aspect ratios. Bottom panel: Quadrupole frequency as a function of $\lambda$ for $\varepsilon_{dd}^0\lambda^{-1/6}=0.5$.Dashed line marks instability as predicted by including exchange interactions (see for instance \cite{Miyakawa2008}).}\label{wq}
\end{figure}

In the collisionless regime the situation is very different. In fact the frequency shift with respect to the ideal gas value $2\omega_\perp$  depends strongly on the geometry of the trap and on the strength of the interaction, as can be seen from the top panels of Fig.~\ref{wq}. In  general we can say that for cigar-shaped clouds, the shift is positive, while for pancake-shaped clouds the shift is always negative.   For $\lambda\sim2$ we find that the shift is very small for all values of $\varepsilon_{dd}$. 
In contrast to the hydrodynamic  regime, in the collisionless regime the frequency shift of the quadrupole mode does not vanish in the limit of large shape deformations $\kappa$ (pancake geometry),  but instead  tends to a value that depends only on $\varepsilon_{dd}$:
\begin{equation}
	\omega_Q^{(\text{CL})}(\kappa\rightarrow\infty)=2\omega_\perp\left(1-\frac{\varepsilon_{dd}}{1+2\varepsilon_{dd}}\right)^{1/2} \ .\label{wqepsilon}
\end{equation}
For very large $\varepsilon_{dd}$ Eq. (\ref{wqepsilon}) approaches the hydrodynamic value  $\sqrt{2}\omega_\perp$.  This is not a surprise since in this limit the kinetic energy contribution to the energy of the dipolar gas is strongly suppressed, the equilibrium profile being the result of the competition between the harmonic potential and the dipolar force. The large $\varepsilon_{dd}$ limit is reminiscent of the zero sound scenario in Fermi liquids where, for large values of the interaction coupling constant, the zero sound velocity approaches the value of first sound.

The bottom panel of Fig.~\ref{wq} shows the dependence of the quadrupole frequency on the trap aspect ratio, for a given value of the dimensionless parameter
\begin{equation}
	\varepsilon_{dd}^0\lambda^{-1/6}=\frac{2^{12}6^{1/6}}{3^4 35 \pi^2}d^2N^{1/6}\left(\frac{m^3\omega_\perp}{\hbar^5}\right)^{1/2}.
\end{equation}
We have identified this parameter since, for a given value of $\omega_\perp$,  it  is independent of the geometry of the gas ($\lambda$ or $\kappa$) or its density. The value of the 
 numerical factor in front of $d^2$ is $0.1973$. To give an order of magnitude, using the experimental parameters of Ref.~\cite{DeMiranda2011}, that is $N=2200$  $^{40}$K$^{87}$Rb molecules in a trap with $\omega_\perp=36\times2\pi$ s$^{-1}$, the condition $\varepsilon_{dd}^0\lambda^{-1/6}=0.5$ corresponds to an electric dipole moment of $0.74$ Debye. We see from the figure that the frequencies in the hydrodynamic and collisionless regimes are sensibly different. They can be used therefore to study the transition (or crossover) between the two regimes. It can take place in two different scenarios: when the gas enters the superfluid phase, or when the gas goes from the collisionless regime to the collisional regime, where hydrodynamic equations are imposed by a high collision rate.

 For very pancake systems, superfluidity cannot take place because the dipole interaction is mainly repulsive. However, when we decrease the value of $\omega_z$ the attractive part of the dipolar interaction can give rise to pairing and superfluidity if temperature is sufficiently low \cite{Baranov2004}. The radial quadrupole mode might then become a good test for  the  transition. For highly pancake traps the frequency should correspond to the collisionless  prediction and hence be given by the upper (red) line in Fig.~\ref{wq} (bottom panel). As the trap aspect ratio is decreased, at some point the attractive dipolar interactions will induce the gas to enter the superfluid phase, and the quadrupole frequency should drop  to its hydrodynamic value, given by the lower line (black) in the figure. To give an order of magnitude, following Ref.~\cite{Baranov2004} for a Fermi gas containing $N=10^6$ atoms or molecules for which $\varepsilon_{dd}^0=0.5$, the transition to the superfluid phase should occur at $\lambda_{cr}\sim 2$: for $\lambda > 2$ the gas would be in the normal phase, while for $\lambda < 4$ it should be a BCS superfluid.
 
Alternatively, in the normal phase at finite $T$, the study of the quadrupole oscillation and the transition from the collisionless to the hydrodynamic regime could provide an important test of the role of the collisions caused by the dipolar interaction. 
Indeed, if $\tau$ is the typical collision time due to the dipolar interaction and $\omega$ the colective mode frequency one gets the collisionless (hydrodynamics) result provided $\omega \tau \gg 1$ ($\omega \tau \ll 1$). If we estimate $\tau$ at low temperature within the Born approximation for the scattering amplitude (see e.g \cite{BaymPethick}), we get 
\be
\frac{1}{\tau}\sim\varepsilon_{dd}^2{\epsilon_F\over\hbar}\left({T\over T_F}\right)^2=\varepsilon_{dd}^2\frac{\omega_\perp}{(6 N\lambda)^{1/3}}\left({k_B T\over\hbar\omega_\perp}\right)^2,
\ee
where, in the last equality we replace $\epsilon_F$ (and $T_F$) with its non-interacting value $\epsilon_F^0=\hbar\omega_\perp(6 N \lambda )^{1/3}$. Since the frequency of the quadrupole oscillation is of order $\omega_\perp$, the condition for being in the collisionless regime is then equivalent to requiring $T\ll T_F/(\varepsilon_{dd}(6N\lambda)^{1/6})$.

\section{Virial theorem and the measurement of the dipolar interaction}
\label{sec:virial}

In Sect. II the radial virial relationship Eq. (\ref{virial}) has been used to calculate the quadrupole mode frequency in the collisionless regime. Such expression 
can be obtained by considering a radial scaling transformation of the form $x\to e^c x$, $y\to e^cy$ applied  to the ground state, with the proper normalization constraint (e.g., the density changes as $n \to e^{-2c}n(e^cx,e^cy,z)$) and by imposing that the total energy (\ref{energy}) vanishes at first order in $c$.
A similar virial relation can be found by considering a scaling transformation along the $z$-th axis and one finds:
 \begin{equation}
2E_{\text{kin},z} - 2E_{\text{ho},z} + E_{\text{dip}} +\frac{1}{4}N\epsilon_F\varepsilon_{dd}h(\kappa)=0\label{virialz}
\end{equation}
By summing the radial and the longitudinal virial expressions Eqs. (\ref{virial}) and (\ref{virialz}) one recovers the more familiar virial expression $2 E_{\text{kin}}-2E_{\text{ho}}+3E_{\text{dip}}=0$ (see, e.g., \cite{collFermi, martavirial}).

The virial expressions   play an important role also in the dynamics of the system following the sudden switching off of the dipolar interaction. Let us suppose that the system, initially at equilibrium in the presence of the dipolar interaction, is perturbed at $t=0$ by suddenly setting $v_{dd}=0$ (in the case of heteronuclear molecules it is enough to set the external electric field equal to zero). For $t>0$ the dynamics of the system is governed by the  single particle  Hamiltonian  $H_0=p^2/2m + 
V_{ho}({\bf r})$. For example the average radial square radius will evolve in time according to the law
\begin{eqnarray}
	{d^2 \over dt^2}\langle r^2_\perp\rangle& =&{4\over m}\left(\left\langle{p^2_\perp\over 2m}\right\rangle -\left\langle{m\omega^2_\perp r^2_\perp\over 2}\right\rangle\right)=\nonumber\\
	&= &{4\over m}E_\perp^0  -{4\over m}\langle m\omega^2_\perp r^2_\perp\rangle
	 \label{measure1}
\end{eqnarray}
where, in the last identity, we have introduced the energy $E_\perp^0=\langle p^2_\perp/2m\rangle +\langle m\omega^2_\perp r^2_\perp/2\rangle$ along  the radial direction. This quantity is conserved in time (for $t>0$), due to the absence of the dipolar interaction. After some straightforward algebra it is easy to find the following time dependence for the radial oscillator energy:
 \begin{eqnarray}
	&E_{\text{ho},\perp}(t)  = E_{\text{ho},\perp}(t=0)\nonumber\\
	&+\left[ E_{\text{kin},\perp}(t=0)- E_{\text{ho},\perp}(t=0)\right] \sin^2(\omega_\perp t)\label{measure2}
\end{eqnarray}
Thanks to the virial theorem (\ref{virial}) the quantity in the square brackets in front of $\sin^2(\omega_\perp t)$ can be directly expressed in terms of the dipolar contribution calculated before switching off the dipolar force.  A similar relationship holds also for the time evolution of the oscillator energy along the $z$-th direction -- same Eq. (\ref{measure2}) with $\perp\rightarrow z$ -- so that, combining the measurements of the time evolution of the radii along the radial and axial directions (and hence of $E_{\text{ho},\perp}(t)$ and $E_{\text{ho},z}(t)$ respectively) one can eventually determine experimentally the value of the dipole energy as well as the function $h(\kappa)$ \cite{dalibard}.

\section{Conclusions}

In this paper we have studied the effect of the dipolar interaction on the surface quadrupole mode in a trapped Fermi gas in both the collisionless and the  (superfluid or collisional) hydrodynamic regime. The predicted values depend on the strength of the interaction and the cloud aspect ratio.
We have in particular found that:

(i) in the hydrodynamic regime the frequency is affected weakly and is always smaller than the ``standard" result $\sqrt{2}\omega_\perp$ which is recovered for a strongly deformed cloud, i.e., when the anisotropic nature of the dipole-dipole potential is irrelevant; 

(ii) in the collisionless regime the frequency of the quadrupole mode can be strongly affected by the dipolar interaction and it is smaller (larger) than the free gas case $2\omega_\perp$ for a pancake (cigar) shape trap.

The difference between the frequencies in the two regimes is large enough to be experimentally resolved.

It is important to notice that the gas can be in the hydrodynamic regime either because of superfluidity or because of collisions. We provided an estimate for the temperature in order the system be collisional for the qudrupole mode. 
 
The virial theorem, which has been employed to derive the quadrupole mode frequency in the collisionless regime, has been also proved to be useful to measure the dipolar energy of the system via a sudden switch-off of the dipolar interaction itself.

\section*{Acknowledgement}

We acknowledge P. Pedri for stimulating discussions. Useful discussions with M. Klawunn, N. Matveeva and C. Menotti are also acknowledged. This work has been supported by ERC through the QGBE grant.
M. A. has been supported by CUR of DIUE (Generalitat de Catalunya) and the European Social Fund. M. A. thanks the group at the BEC center and the University of Trento for their kind hospitality, nice discussions and great fun.

\appendix 

\section{Anisotropy functions $f(\kappa)$, $g(\kappa)$ and $h(\kappa)$}

\paragraph{The function $f(\kappa)$}

\begin{figure}[h!]
 \epsfig{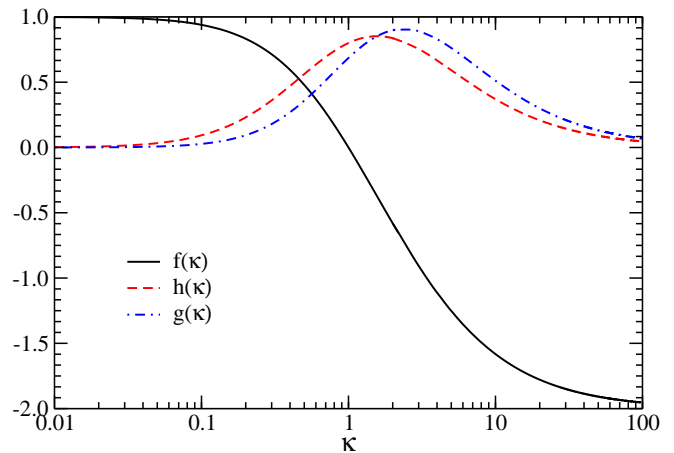}
\caption{Anisotropy function $f(\kappa)$ and its derivatives, $h(\kappa)$ and $g(\kappa)$, that appear in the expressions of the quadrupole frequency, as a function of $\kappa$.}\label{fgh}
\end{figure}

The function $f(\kappa)$ reflects the anisotropy of dipolar interactions both for bosonic and fermionic systems. For a cylindrically symmetric system it can be written as \cite{ODell2004}
\begin{equation}
	f(\kappa)= \frac{1+2	\kappa^2}{1-\kappa^2}-\frac{3\kappa^2}{(1-\kappa^2)^{3/2}}\tanh^{-1}\sqrt{1-\kappa^2}\ .\label{fkappa}
\end{equation}
If the system has no cylindrical symmetry, the anisotropy function depends on the aspect ratios $\kappa_x=R_x/R_z$ and $\kappa_y=R_y/R_z$ and has the form  \cite{Giovanazzi2006}
\begin{equation}
	f(\kappa_x,\kappa_y)=1+3\kappa_x\kappa_y\frac{E(\varphi\backslash\alpha)-F(\varphi\backslash\alpha)}{(1-\kappa_y^2)\sqrt{1-\kappa_x^2}} \ ,
\end{equation}
where  $F(\varphi\backslash\alpha)$ and $E(\varphi\backslash\alpha)$ are the incomplete elliptic integrals of first and second kind \cite{Abramowitz} and their arguments are given by:
\begin{align}
	& \sin^2\alpha = \frac{1 - \kappa_y^2}{1 - \kappa_x^2}\\
	& \sin\varphi = \sqrt{1-\kappa_x^2} \ .
\end{align}
Much information about this function can be found in Ref.~\cite{Giovanazzi2006}. It can be seen that when the two arguments are equal, $\kappa_x=\kappa_y$, one recovers expression (\ref{fkappa}). Figure \ref{fgh} shows the behavior of $f(\kappa)$ with respect to its argument. Note that it vanishes for $\kappa=1$, implying thus that the dipolar energy vanishes for a spherical dipolar gas. It is also interesting to notice that in the limit $\kappa\rightarrow\infty$ the function tends to a constant value, $f(\kappa)\rightarrow -2$, and the dipolar energy depends on the Thomas-Fermi radii in the same way as does a contact interaction. 
This is the underlying reason why the quadrupole frequency is not affected by dipolar interactions in the HD regime, when the gas approaches the 2D limit.

\paragraph{The function $g(\kappa)$}

In the expression of the quadrupole frequency, in both regimes, there appears the function $g(\kappa)$ defined by
\begin{align}
	&g(\kappa) = \frac{\partial^2 f(e^b\kappa,e^{-b}\kappa)}{\partial b^2}\bigg|_{b=0} = \nonumber\\
	&=\frac{3\kappa^2}{2}\frac{\sqrt{1-\kappa^2}(2+13\kappa^2)-\kappa^2(12+3\kappa^2)\tanh^{-1}\sqrt{1-\kappa^2}}{(1-\kappa^2)^{7/2}}
\end{align}
It is shown in Fig.~\ref{fgh}. It vanishes for $\kappa\rightarrow0$ and $\kappa\rightarrow\infty$ and has a maximum at $\kappa\simeq 2.37$.

\paragraph{The function $h(\kappa)$}

In the radial and longitudinal virial expressions we have introduced the function $h(\kappa)$ defined by
\begin{align}
	&h(\kappa) = -\kappa\frac{d f(\kappa)}{d\kappa}   \nonumber\\
	&=\frac{3\kappa^2}{(1-\kappa^2)^{5/2}}\left[-3\sqrt{1-\kappa^2} + (2+\kappa^2)\tanh^{-1}\sqrt{1-\kappa^2}\right]
\end{align}
The behavior of the function $h(\kappa)$  is very similar to the one of $g(\kappa)$. It  vanishes both for $\kappa\rightarrow0$ and $\kappa\rightarrow\infty$ and has a maximum at $\kappa\simeq1.56$ (see Fig.~\ref{fgh}).

\end{document}